\documentclass[twocolumn,aps,prl,amsmath,floatfix]{revtex4}
\usepackage{graphicx}
\begin{document}
\title{Comment on ``Orbital-selective Mott transitions in
                    two-band Hubbard models''}
\author{A.~Liebsch} 
\affiliation{Institut f\"ur Festk\"orperforschung, 
             Forschungszentrum J\"ulich, 
             52425 J\"ulich, Germany}
\begin{abstract} 
A recent paper by Bl\"umer {\it et al.} [cond-mat/0609758] again 
criticizes earlier QMC/DMFT results by Liebsch 
[Phys. Rev. B {\bf 70}, 165103 (2004)]. 
This criticism is shown to be unfounded. 
\end{abstract}
\maketitle
In Ref.~1 Bl\"umer {\it et al.}~continue to criticize 
earlier QMC/DMFT calculations by Liebsch \cite{prb70} for the 
non-isotropic two-band Hubbard model.
Now it is claimed: ``We quantify numerical errors in earlier 
QMC data which had obscured the second transition'' and:
``The second transition is lost in the noise of earlier data [2]
with errors exceeding  100 \% at both transitions''.

We point out that Ref.~1 once again does not provide any 
comparisons of self-energies or spectral distributions [3], nor
does it refer to recent work [4,5] which confirms the results 
of Ref.~2. A direct comparison demonstrates, as we 
show here, that both QMC calculations are in good agreement 
and that the above claims are unfounded. 

\begin{figure}[b!]  
  \vskip-2mm
  \begin{center}
  \includegraphics[width=5.0cm,height=8cm,angle=-90]{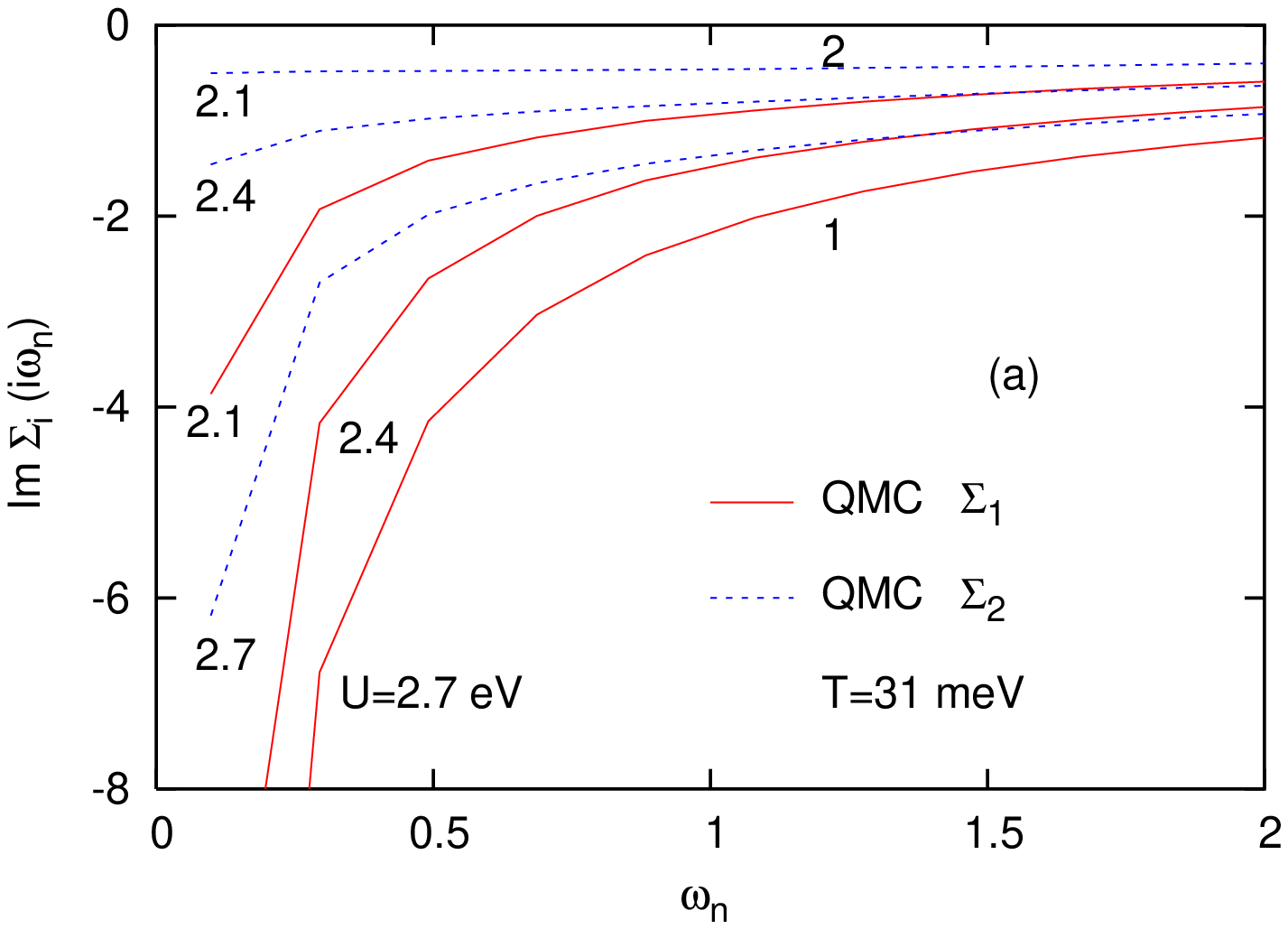}
  \includegraphics[width=5.0cm,height=8cm,angle=-90]{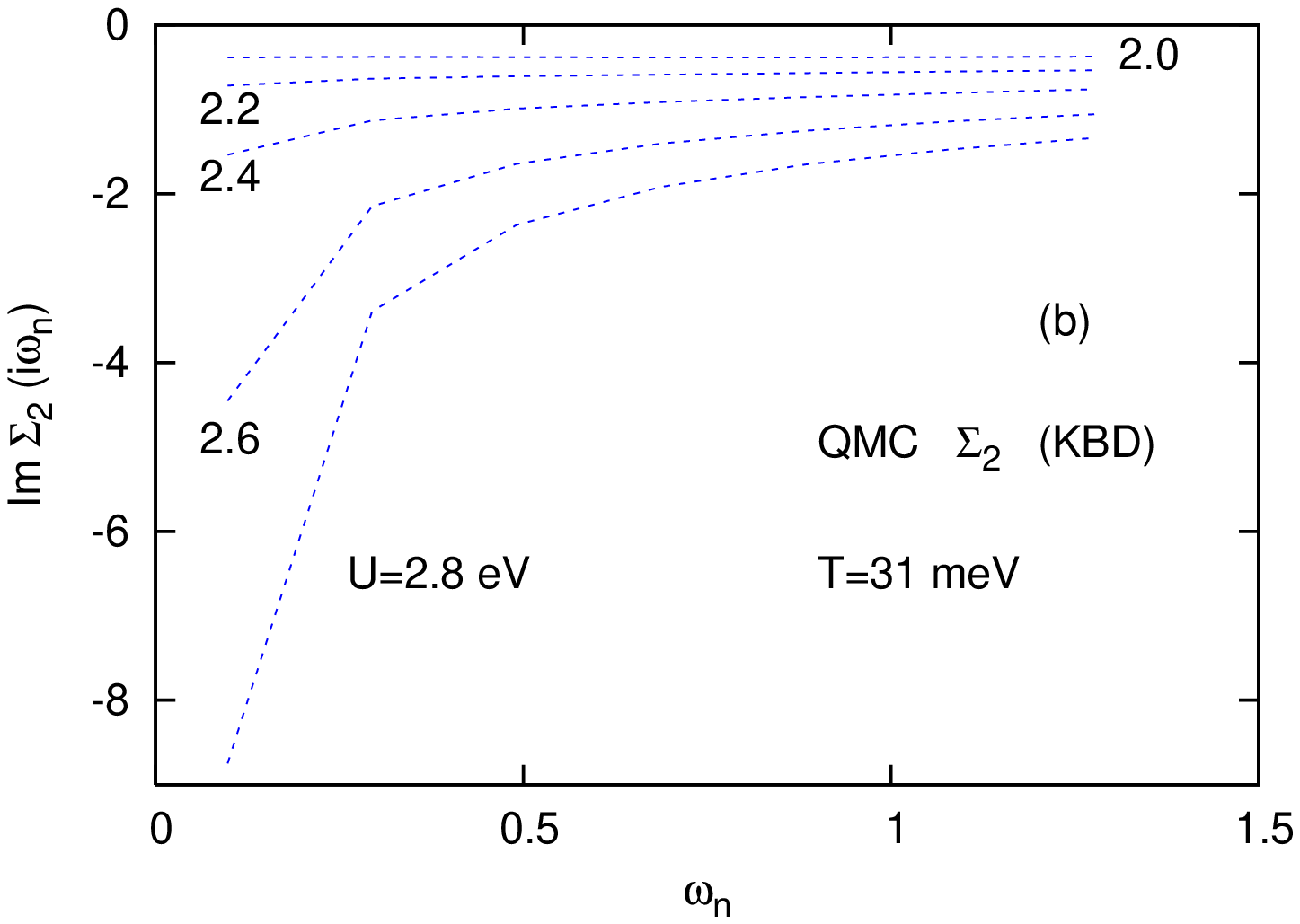}
  \end{center}
  \vskip-3mm
\caption{
(a) QMC subband self-energies $\Sigma_i(i\omega_n)$ 
for different $U$, from Fig.~10 of Ref.~2.
Solid red curves: narrow band; dashed blue curves: wide band.  
(b) QMC self-energy of wide band, adapted from Fig.~2 of Ref.~1 (KBD).
The narrow band is insulating in this range.
}\end{figure}

Fig.~1 shows the comparison of QMC self-energies calculated 
in Ref.~2 with those of Ref.~1. Fig.~2 shows analogous results
obtained via exact diagonalization (ED) and 
numerical renormalization group (NRG) \cite{al+costi}.

\begin{figure}[b!]  
  \vskip-2mm
  \begin{center}
  \includegraphics[width=5.0cm,height=8cm,angle=-90]{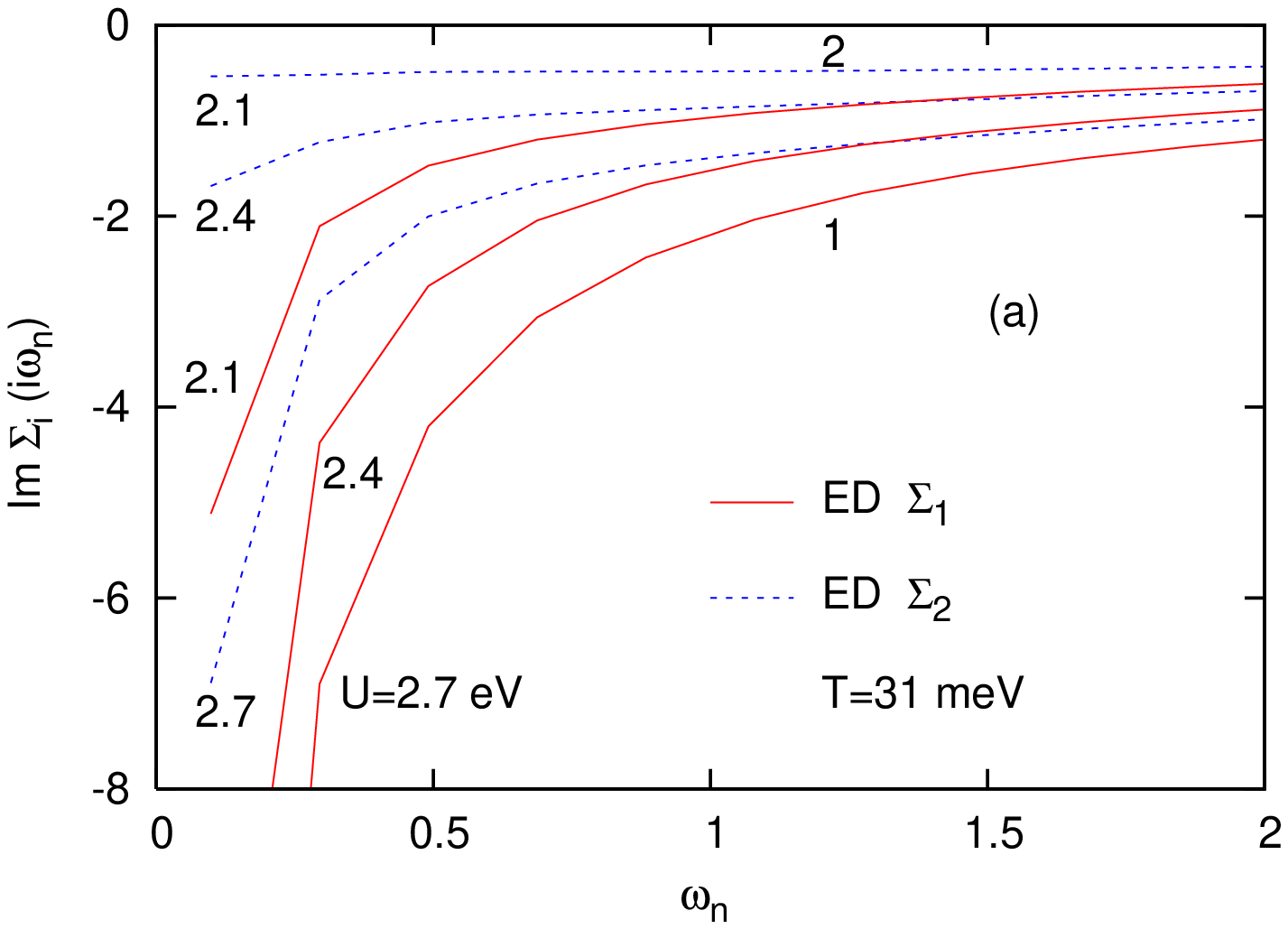}
  \includegraphics[width=5.0cm,height=8cm,angle=-90]{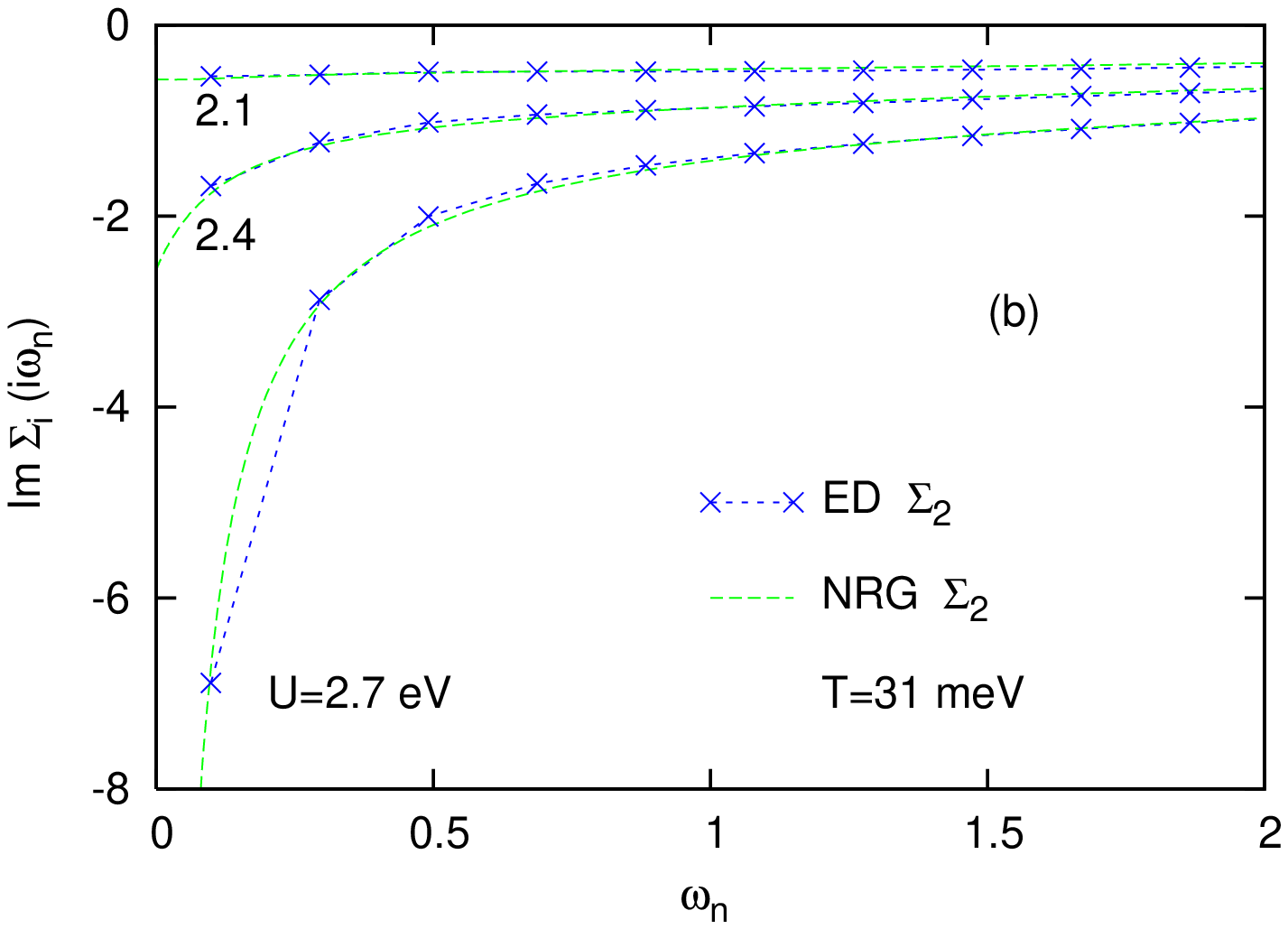}
  \end{center}
  \vskip-3mm
\caption{
(a) ED subband self-energies for the same parameters as in Fig.~1(a).
Solid red curves: narrow band; dashed blue curves: wide band.  
(b) Comparison of ED and NRG self-energies of wide band.
The narrow band is insulating in this range. 
(Both results from Ref.~5).
}\end{figure}

Evidently, all calculations give the same trend: When the 
narrow band becomes insulating, the self-energy of the wide band no
longer exhibits $\sim i\omega_n$ behavior at low frequencies, as  
would be characteristic of a Fermi-liquid. Instead, it shows 
progressive bad-metallic behavior, approaching a finite value in 
the $i\omega_n\rightarrow0$ limit. This value grows with increasing
$U$, until it diverges near 2.7~eV. Ref.~2 states:
``$\Sigma_2(i\omega_n)$ becomes inversely proportional to $\omega_n$ 
at 2.7~eV, i.e., a gap opens up.''

Precisely this behavior is seen in the quasi-particle spectra derived 
in Refs.~1 and 2 (see Fig.~3).
Despite the differences caused by different maximum entropy
fitting parameters, the low-frequency region is in perfect agreement.
Both spectra show that, when the narrow band becomes insulating, the 
wide band reveals a pseudogap which gets progressively deeper with 
increasing $U$, until this band becomes fully insulating near 
2.7~eV.  

\begin{figure}[t!] 
  \begin{center}
  \includegraphics[width=4.1cm,height=3.9cm,angle=-90]{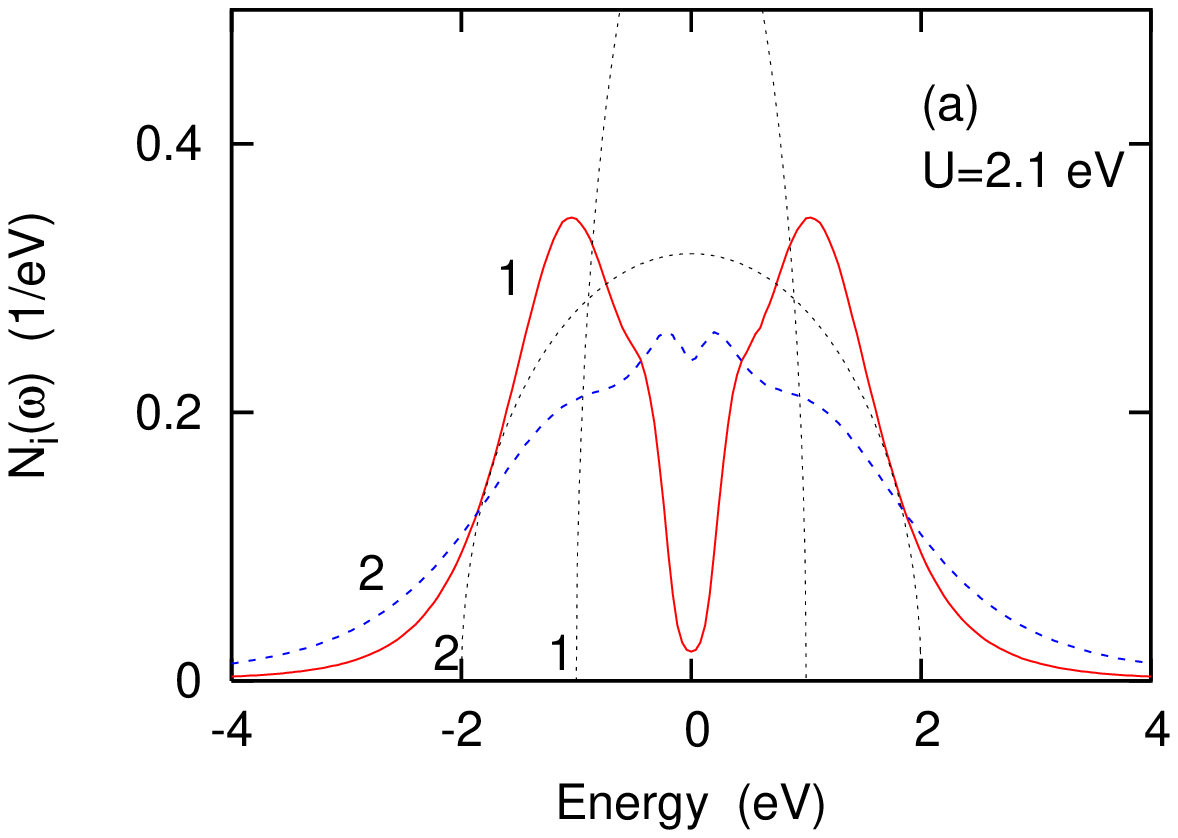}
  \includegraphics[width=4.1cm,height=3.9cm,angle=-90]{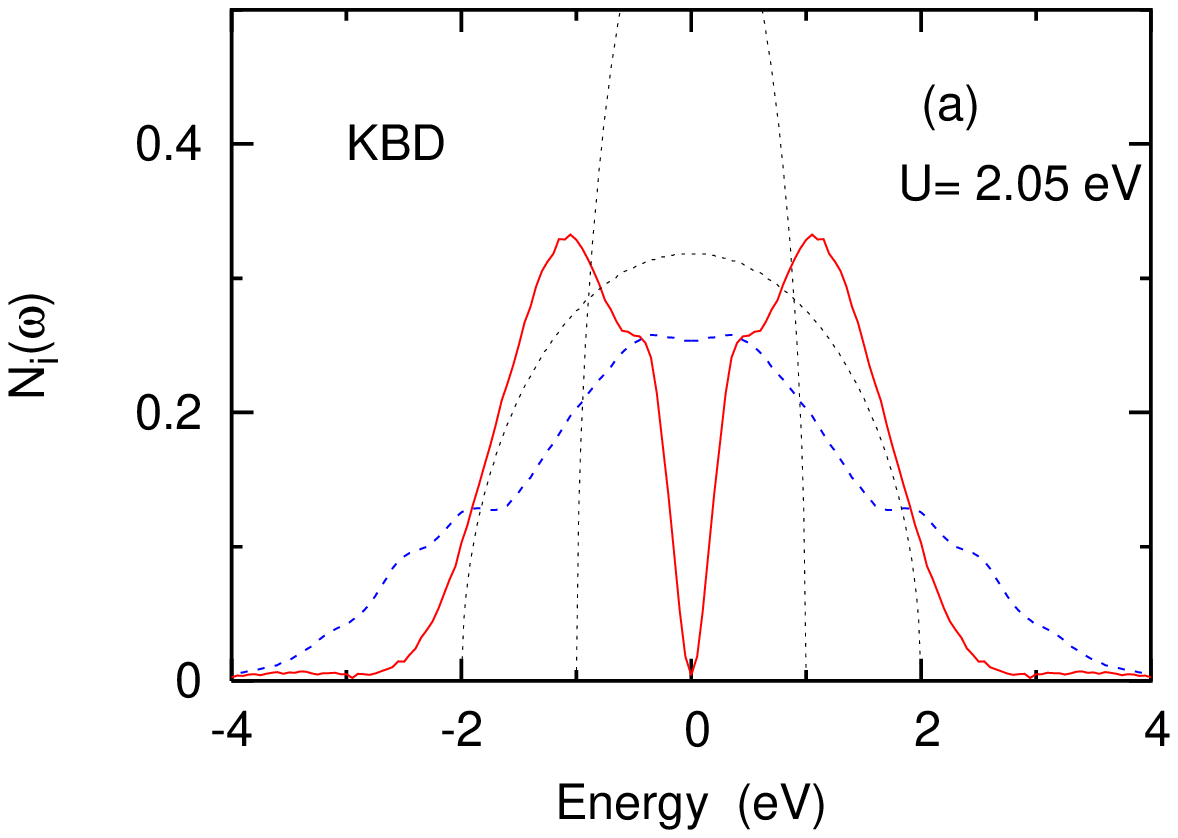}
  \includegraphics[width=4.1cm,height=3.9cm,angle=-90]{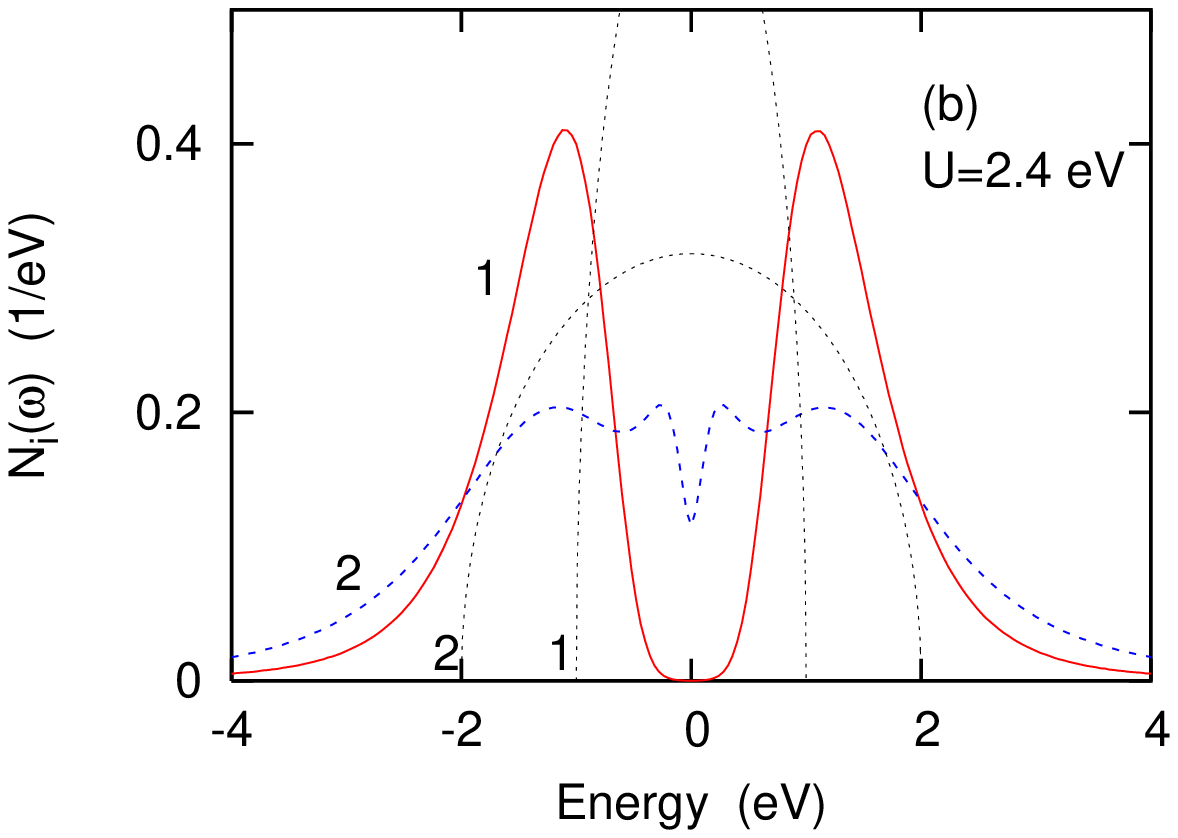}
  \includegraphics[width=4.1cm,height=3.9cm,angle=-90]{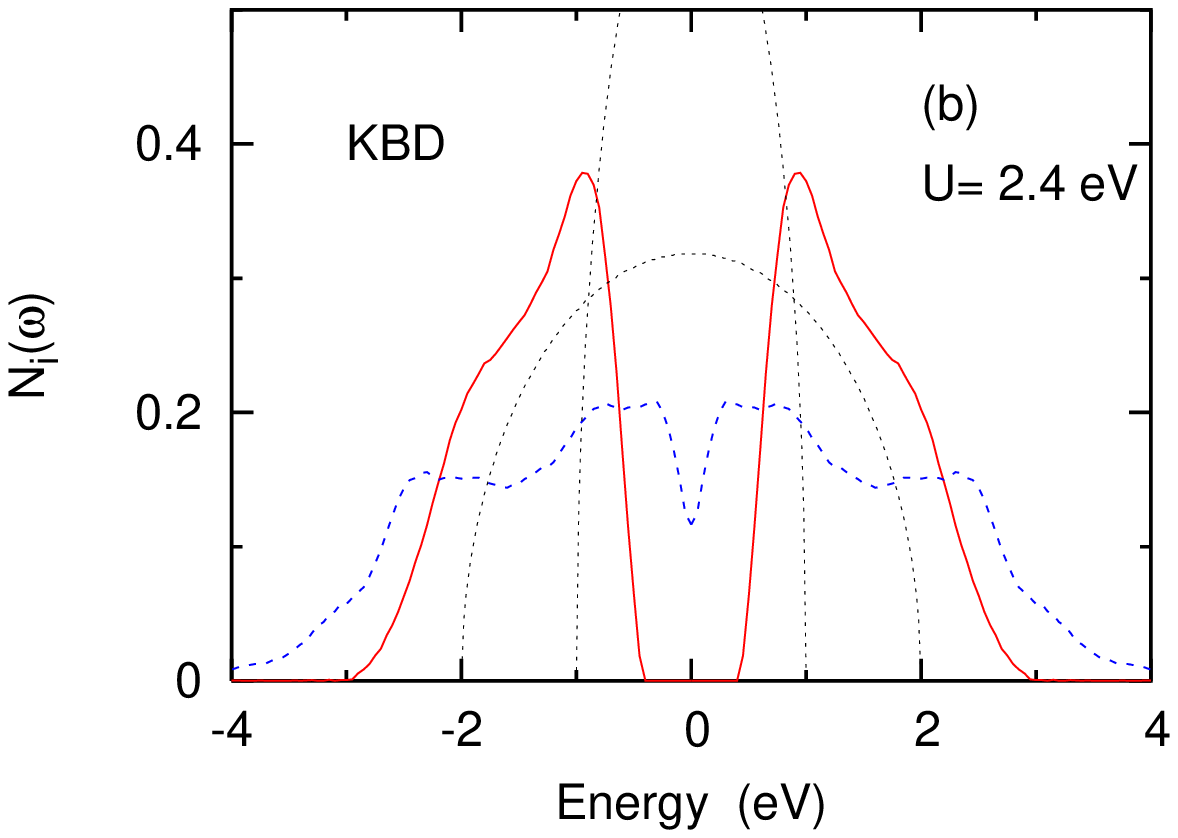}
  \includegraphics[width=4.1cm,height=3.9cm,angle=-90]{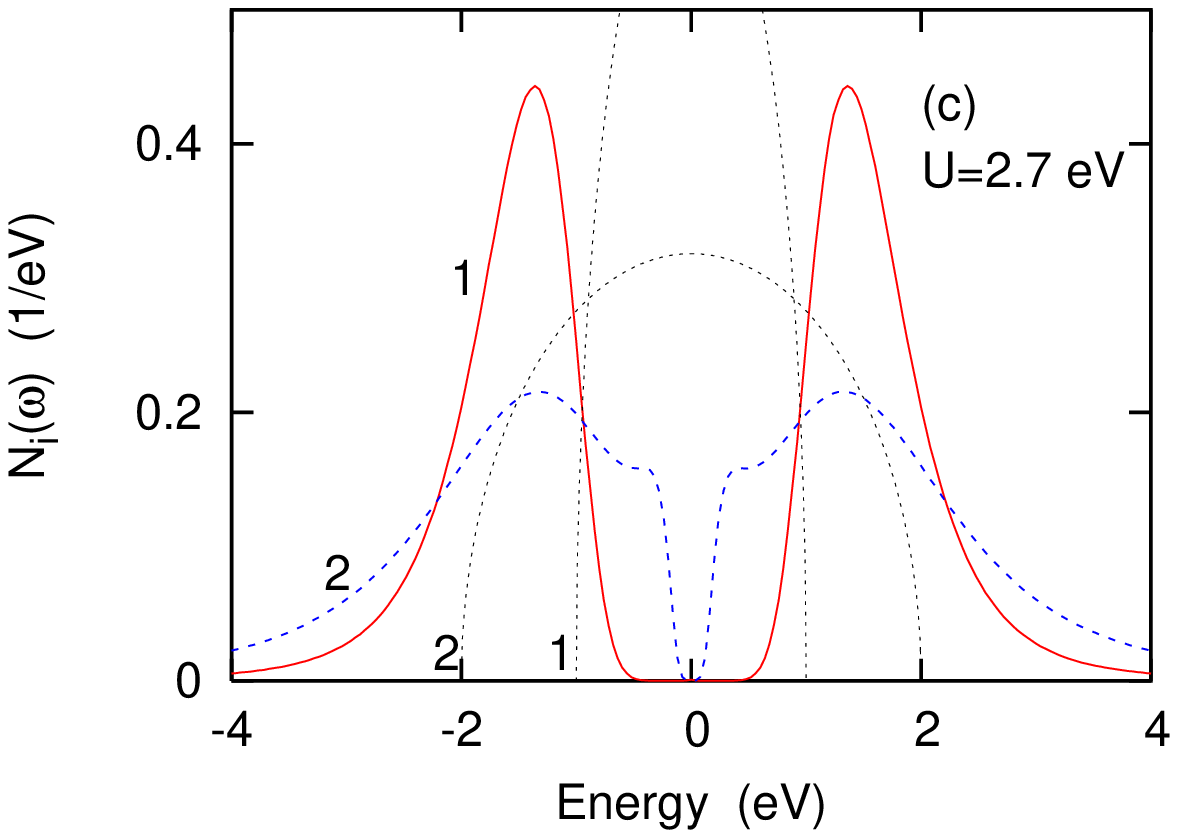}
  \includegraphics[width=4.1cm,height=3.9cm,angle=-90]{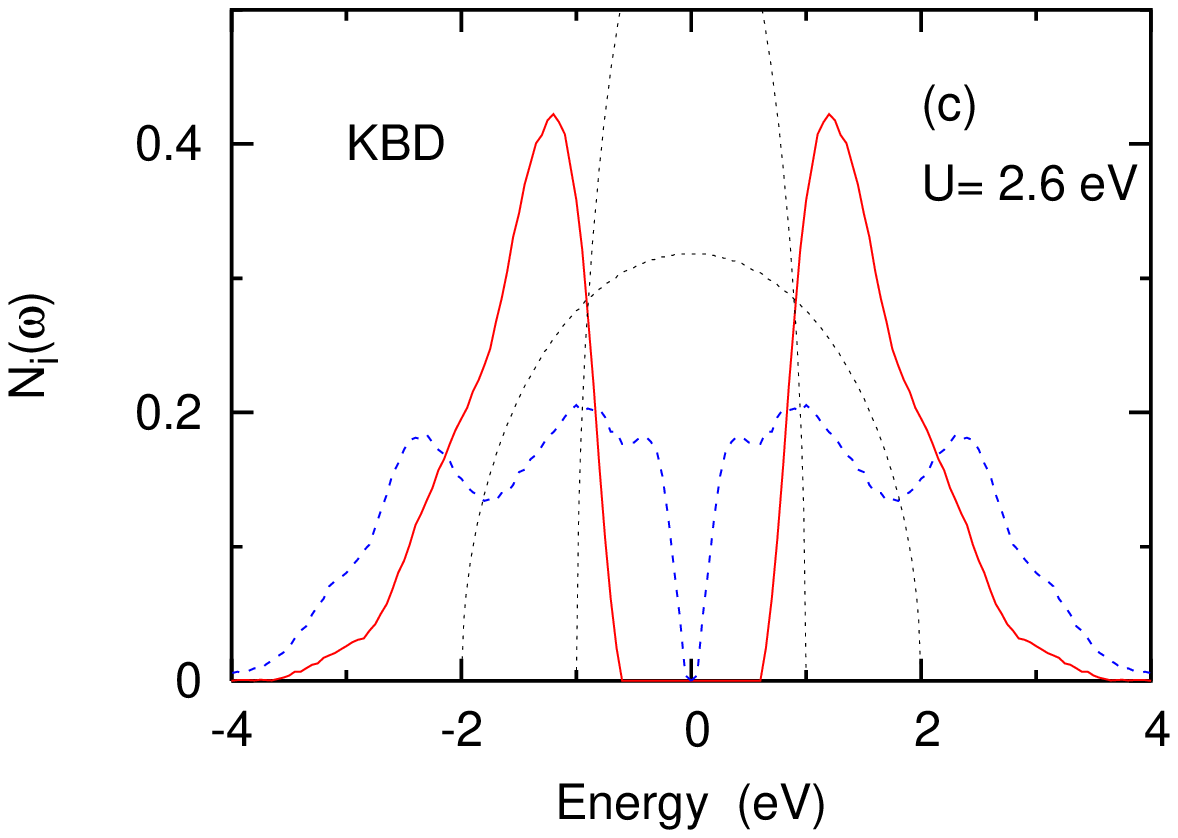}
  \end{center}  
 \vskip-3mm
\caption{
Quasiparticle spectra from Ref.~2 at $T=31$~meV (left panels) 
and from Ref.~1 (KBD) at $T=25$~meV (right panels). 
Solid red curves: narrow band; dashed blue curves: wide band; 
dotted curves: bare densities of states. (See Ref.~3.) 
} \vskip-2mm
\end{figure}

The self-energies and spectral distributions indicate that there 
is good agreement between the QMC results of Refs.~1 and 2. 
Both describe identical physics and are consistent with the ED 
and NRG results. There is no evidence of any sort of disparity. 
Thus, the claims in Ref.~1 have no basis.

Fig.~4 compares $Z_i(U)= 1/[1-{\rm Im}\Sigma_i(i\omega_0)/i\omega_0]$. 
Again, there is good agreement, consistent with the results in Figs.~1--3. 
The main difference is that the Coulomb energies at which subbands 
become insulating are slightly lower in Ref.~1 than in Ref.~2 (see
also Fig.~3).

Although the $Z_i(U)$ are derived from the self-energies in Fig.~1, 
Ref.~1 claims that $\Delta Z_i(U)$ (obtained by subtracting results 
from Refs.~1 and 2) reveals a qualitative difference: 
``second transition lost in noise'', etc.

A proper analysis of $\Delta Z_i(U)$ should, of course, include 
(i) the different error margins resulting from QMC statistical 
uncertainties, number of sweeps and time slices, and 
(ii) the different Coulomb energies at which subbands become insulating,
for instance, as a result of a different $U$ mesh, different number of 
iterations and critical slowing down. These issues are particularly important 
when $\Sigma_i(i\omega_n)$ becomes singular and $Z_i(U)$ becomes small.
Since all of this is ignored in Ref.~1, it is no surprise that the  
agreement seen in Fig.~4 can be turned, at specific points, into 
ficticious disagreement of $\Delta Z_i(U)$  of arbitrary magnitude.

Evidently the criticism in Ref.~1 is based on a fundamentally inadequate 
analysis of $\Delta Z_i(U)$. Moreover, Ref.~1 does 
not provide the reader with direct comparisons of self-energies or 
spectral functions, such as given here in Figs.~1,3 or in Ref.~3,
which demonstrate good agreement. Finally, Ref.~1 ignores that the 
QMC results of Ref.~2 were fully confirmed by ED and NRG calculations [4,5]. 

\begin{figure}[t!]  
  \begin{center}
  \includegraphics[width=5.0cm,height=8cm,angle=-90]{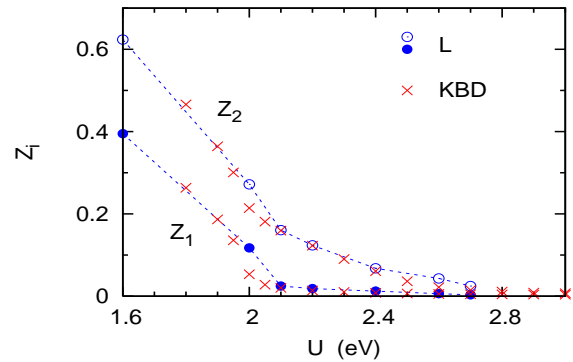}
  \end{center}
  \vskip-3mm
\caption{
Comparison of $Z_i(U)$ derived within QMC for 
$T=31$~meV, Ising exchange with $J=U/4$ \cite{comment}. 
$Z_1$: narrow band; $Z_2$: wide band. 
Solid and open dots (blue): results of Ref.~2 (L); 
crosses (red): results of Ref.~1 (KBD).
}\vskip-3mm
\end{figure}

We conclude that the QMC/DMFT results of Ref.~2 are correct:
The non-isotropic two-band Hubbard 
model with Ising exchange exhibits a first-order Mott transition near 
$U=2.1$~eV when the narrow band becomes insulating, with characteristic 
hysteresis behavior, and  there is no sign of first-order behavior 
when the wide band becomes insulating near $2.7$~eV. 

As also shown by the ED/DMFT calculations in Ref.~4, to obtain sequential 
first-order Mott transitions, it is essential to go beyond Ising exchange 
and include full Hund's coupling.   

\vfill
\end{document}